\documentclass[prb,twocolumn,superscriptaddress,amsmath,amssymb,showpacs]{revtex4}%
\usepackage{graphicx}
\usepackage{dcolumn}
\usepackage{bm}

\begin{document}
\title{Polarization-dependent shift in excitonic Zeeman splitting\\ of self-assembled InAlAs quantum dots}

\author{T.\ Yokoi}
	\email{piyoko@eng.hokudai.ac.jp}
	\affiliation{Department of Applied Physics, Hokkaido University, Sapporo 060-8628, Japan}

\author{S.\ Adachi}
	\affiliation{Department of Applied Physics, Hokkaido University, Sapporo 060-8628, Japan}
	\affiliation{CREST, Japan Science and Technology Agency, Kawaguchi 332-0012, Japan}

\author{H.\ Sasakura}
	\affiliation{Department of Applied Physics, Hokkaido University, Sapporo 060-8628, Japan}
	\affiliation{CREST, Japan Science and Technology Agency, Kawaguchi 332-0012, Japan}

\author{S.\ Muto}
	\affiliation{Department of Applied Physics, Hokkaido University, Sapporo 060-8628, Japan}
	\affiliation{CREST, Japan Science and Technology Agency, Kawaguchi 332-0012, Japan}
	
\author{H.\ Z.\ Song}
	\affiliation{CREST, Japan Science and Technology Agency, Kawaguchi 332-0012, Japan}

\author{T.\ Usuki}
	\affiliation{CREST, Japan Science and Technology Agency, Kawaguchi 332-0012, Japan}

\author{S.\ Hirose}
	\affiliation{Fujitsu LIMITED, 10-1 Morinosato-Wakamiya, Atsugi 243-0197, Japan}



\begin{abstract}
We report the optical spectroscopic results of a single self-assembled In$_{0.75}$Al$_{0.25}$As/Al$_{0.3}$Ga$_{0.7}$As quantum dot. The polarization-dependent shift of the Zeeman splitting in a single InAlAs QD has been observed. The induced Overhauser field is estimated to be $\sim$0.16 T in this InAlAs QD and the magnitude is shown to be controllable by the degree of circular polarization of excitation light. 
\end{abstract}
\pacs{78.67.Hc, 71.70.Ej}
\maketitle

\section{Introduction}
Semiconductor self-assembled quantum dots (QDs) exhibit a variety of confinement-related optical and electronic properties useful for opto-electronic device applications such as QD lasers and detectors. Especially, the broad effort is now under way to develop new techniques for controlling spin degrees of freedom in QDs. These efforts are stimulated in part by some proposals to use the spin systems as quantum bits in quantum information processing~\cite{Loss98,Muto02,SSQC02}. While the rapid spin relaxation in solid-state surroundings was regarded as the main obstacle for realization of coherent control experiments, the exciton spin relaxation is getting recognized to far exceed the exciton lifetime and lasts up to several nanoseconds if excitons are excited and detected resonantly~\cite{Watanuki04}.
As the next problem, there is the influence of the nuclear-spin-induced magnetic field on the electronic energy states. Since the relaxation time of nuclear spin is extremely long, the induced electronic energy shifts due to nuclear spin polarization will generate errors in quantum gate operation using Zeeman splitting~\cite{Sasakura04c}. Also, we have proposed to use the nuclear magnetic field to realize the qubit conversion from photon qubit to spin qubit~\cite{Muto04}.
The magnitude of nuclear-spin-induced magnetic field and its controllability in a self-assembled QD should be  studied experimentally. \\
\ \ In this work, we report the magnetic field studies of a self-assembled single InAlAs/AlGaAs QD. The polarization-dependent energy shift of excitonic emission, that is well known as Overhauser shift, is clearly observed and the magnitude is shown to be controllable. There is a report on the observation~\cite{Brown96,Gammon01} in a naturally formed QD using monolayer-fluctuation of a quantum well. However, this is the first observation in a self-assembled QD, which is suitable for formation of vertically coupled QDs, to the best of our knowledge. Also, this work gives valuable information about the red-emitting InAlAs/AlGaAs QDs where only a few works have been reported ~\cite{PDWang96,Hinzer01} while most of single QD measurements have been performed so far in combinations such as InAs/GaAs\cite{Ortner03} and InGaAs/GaAs\cite{Bayer02} with emission in infrared. 

\section{sample and ensemble photoluminescence}
\begin{figure}[h]
  \begin{center}
    \includegraphics[width=240pt]{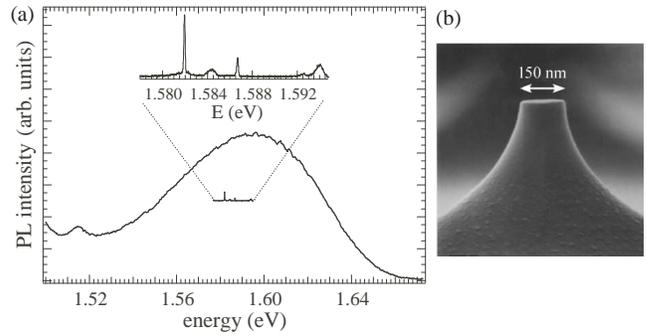}
  \end{center}
  \caption{(a) ensemble PL spectra and single QD emissions (inset) from InAlAs QDs.  (b) Mesa structurer.}
    \label{macroPL}
\end{figure}

The QD samples grown by molecular beam epitaxy have two QDs layers (In$_{0.75}$Al$_{0.25}$As and In$_{0.7}$Ga$_{0.3}$As) separated by a 11-nm-thick Al$_{0.3}$Ga$_{0.7}$As layer. The QDs are formed using the spontaneous island formation in Stranski-Krastanow growth mode during the epitaxy of strained InAlAs (or InGaAs) on AlGaAs layers, grown on CrO-doped (100) GaAs substrates.  A GaAs cap terminates the heterostructure. 
In this study, we concentrate on the single QD emission from InAlAs QDs of this sample. The detail of this sample is seen in refs. \onlinecite{Sasakura04a} and \onlinecite{Sasakura04b}. 

Figure \ref{macroPL} (a) shows the time-integrated photoluminescence (PL) spectra at 10 K from the  excitation spot with the diameter of $\sim 150$ $\mu$ m. The excitation has been carried out with HeNe laser on Al$_{0.3}$Ga$_{0.7}$As barrier. At the lowest excitation intensity, the peak centered around $\sim 1.59$ eV for InAlAs QDs is observed. The PL have the linewidth of the  $\sim 120$ meV due to inhomogeneous QD size distribution. The WL emission (not shown) was observed at 1.689 eV for larger excitation intensity.  

For a detailed understanding of physics governing the properties of the QDs system, it is helpful to go beyond measurements of ensemble-averaged of QDs.
Advances in QD fabrication and in lithographic processing techniques have enabled elegant spectroscopic studies of the optical and electronic properties of a single QD, revealing details of the system typically obscured by ensemble-averaged measurements. 
To isolate a single QD, small mesa structures were fabricated by electron-beam lithography and wet chemical etching as shown in \ref{macroPL} (b). The typical top lateral size of the mesa structure is 150 nm. The QD density is estimated as $\sim$5$\times$10$^{10}$ cm$^{-2}$. A mesa contains several QDs on the average and some mesas have one or a few QDs, from which well-separated sharp emissions appear by conventional far-field spectroscopy as shown in the inset of Fig. \ref{macroPL} (a). For this single QD spectroscopy, the sample was held in a LHe-cryostat and was kept at 4.2 K. The QD emissions were dispersed by a triple grating spectrometer ($f$=0.64 m) and was detected with a LN$_2$-cooled Si-charge coupled devices (CCD) camera. The system resolution was 25 $\mu$eV and the spectral resolution that determines the resonance energies was on the order of 5 $\mu$eV. Typical accumulation times are 5 seconds. The magnetic field up to 5 T was applied to the sample along the growth direction. The polarization of the PL emissions was analyzed with a quarter-wave plate and a linear polarizer in front of the spectrometer.

\section{Zeeman interaction}
\begin{figure}[hp]
  \begin{center}
    \includegraphics[width=240pt]{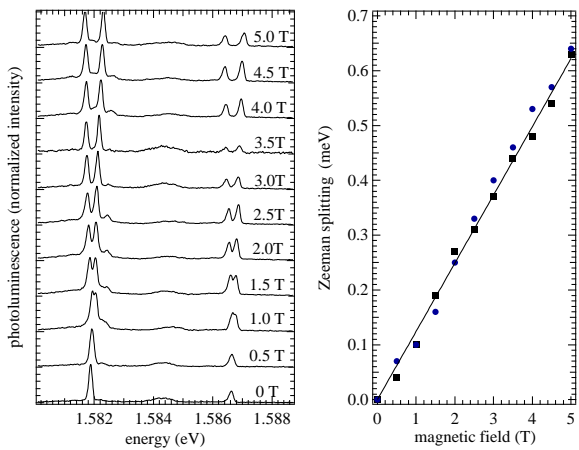}
  \end{center}
  \caption{(a) Single InAlAs-QD PL spectra ($d$=11 nm) recorded at different magnetic fields. The spectra are normalized by the magnitude of the higher energy exciton around 1.5818 eV. (b) Magnetic-field dependence of the exciton splitting of two exciton lines in InAlAs QDs shown in (a). Solid squares (for higher energy exciton) and circles (for lower energy exciton) are experimental data and a line is a fitting result using the forms obtained from diagonalizing the exciton fine-structure Hamiltonian.}  \label{Zeeman}
\end{figure}

Figure \ref{Zeeman}(a) shows the PL spectra from the lowest exciton state of a InAlAs QD with varying the magnetic field up to 5 T at 4.2 K. The magnetic field was aligned parallel to the heterostructure growth direction $z$ and the sample was excited in Faraday configuration. The excitation was linearly-polarized and its power was decreased to a level where biexciton and  excited states disappear in the spectra. In the energy range of the figure, the exciton recombination at $B$=0 T is located at 1.5818 eV. Under low magnetic field ($\le $ 1.5 T) where Zeeman splitting of the exciton line is not clearly observed, the emission has been analyzed with respect to its circular polarization.  
The zero-field emission has the full-width of $\sim$90 $\mu$eV (FWHM) and the linewidth varies from dot to dot within 30-200 $\mu$eV. 
While a very small energy splitting ($\sim$ 14 $\mu$eV) was found at $B$=0 T depending on the observed QDs, they showed no significant linear polarization ~\cite{Bayer02}. In $B\neq$0, the emissions split into a doublet due to Zeeman interaction of the exciton spin with the magnetic field. The low-energy part of the spectrum is $\sigma_+$-polarized and the high-energy one is $\sigma_-$-polarized.   
In this QD, a single emission is due to the recombination of the degenerate $m_j = \pm 1$ excitons. Applying a magnetic field results in a splitting of $m_j = +1$ and $m_j = -1$ states.
As shown in the figure, a diamagnetic shift $\Delta_{diamag}$ to higher energies, which is the energy shift of center of the exciton doublet, is expressed well as $\Delta_{diamag} = \alpha B^2$. For the QD of figure \ref{Zeeman}, the coefficient $\alpha$ is found to be very small (3.4 $\mu$eV/T$^2$). The value of $\alpha$ has been reported so far as 8.6$\pm$0.9 $\mu$eV/T$^2$ for In$_{0.55}$Al$_{0.45}$As/Al$_{0.35}$Ga$_{0.65}$As QDs~\cite{PDWang96} and 0.8$\pm$0.3 $\mu$eV/T$^2$ for In$_{0.64}$Al$_{0.36}$As/Al$_{0.33}$Ga$_{0.67}$As QDs.
Since the diamagnetic shift is proportional to the squared average of lateral extension of exciton wavefunction $\langle r_{\rm x}^2 \rangle$, the small $\alpha$ indicates the strong confinement and it is natural to observe the different $\alpha$ depending on the lateral dot size, especially, for self-assembled of the QDs. Actually, for this sample, some of the QDs exhibited larger diamagnetic shifts.

In Fig. \ref{Zeeman}(b), Zeeman splitting is plotted with the external magnetic field. 
 The exciton energies in Faraday configuration are given as the following Hamiltonian using the exciton states $\displaystyle \left| { m_j} \right\rangle=
\left| { + 1} \right\rangle ,\left| { - 1} \right\rangle ,\left| { + 2} \right\rangle ,\left| { - 2} \right\rangle 
$ as basis;
\begin{eqnarray}
  H &=& H_{exchange}  + H_{Zeeman}  \nonumber\\ 
   &=& \frac{1}{2}\left( {\begin{array}{*{20}c}
   {\begin{array}{*{20}c}
   {\delta _0 } & {\delta _b }  \\
   {\delta _b } & {\delta _0 }  \\
\end{array}} & 0  \\
   0 & {\begin{array}{*{20}c}
   { - \delta _0 } & {\delta _d }  \\
   {\delta _d } & { - \delta _0 }  \\
\end{array}}  \\
\end{array}} \right) \nonumber\\ 
   &+& \frac{{\mu _B B }}{2}\left( {\begin{array}{*{20}c}
   {  g_{b{\rm x}} } & 0 & 0 & 0  \\
   0 & {-g_{b{\rm x}} } & 0 & 0  \\
   0 & 0 & {-g_{d{\rm x}} } & 0  \\
   0 & 0 & 0 & {  g_{d{\rm x}} }  \\
\end{array}} \right) .
\end{eqnarray}
In the above equation, $\delta_0$, $\delta_b$, $\delta_d$ are exchange energy between bright ($\left| { \pm 1} \right\rangle$) and dark ($\left| { \pm 2} \right\rangle$) excitons, splitting energy between bright excitons, and that between dark excitons, respectively. The $\mu_B$ is Bohr magneton and, $g_{b{\rm x}}$ ($g_{d{\rm x}}$) is the g-factor of bright (dark) exciton, given by $g_{e}^{z}+g_{h}^{z}$ ($g_{e}^{z}-g_{h}^{z}$) using electron and hole $g$-factors in growth direction.

By the fitting using the forms obtained from diagonalizing the above exciton Hamiltonian, we obtain an exciton $g$-factor $g_{b{\rm x}}=2.10 \pm 0.03$. A number of single InAlAs QDs were studied in this sample, and the Zeeman splitting changed slightly from dot to dot within $0.1$ meV. This also held for the sample with $d=20$ nm.

\section{Polarization-dependent shift of Zeeman splitting}
In this section, the results in the excitation of circular polarization are reported. In the case of circularly-polarized excitation, the formation of the nuclear spin polarization via hyperfine interaction with spin-polarized electron is expected. 
As already mentioned, the polarization-dependent shift (Overhauser shift) of a single QD emission was observed in a natural GaAs/AlGaAs QD where excitons are trapped in the monolayer fluctuation of quantum well width\cite{Brown96,Gammon01}, but for self-assembled QD the energy shift has not been observed so far.
\begin{figure}[ht]
  \begin{center}
    \includegraphics[width=200pt]{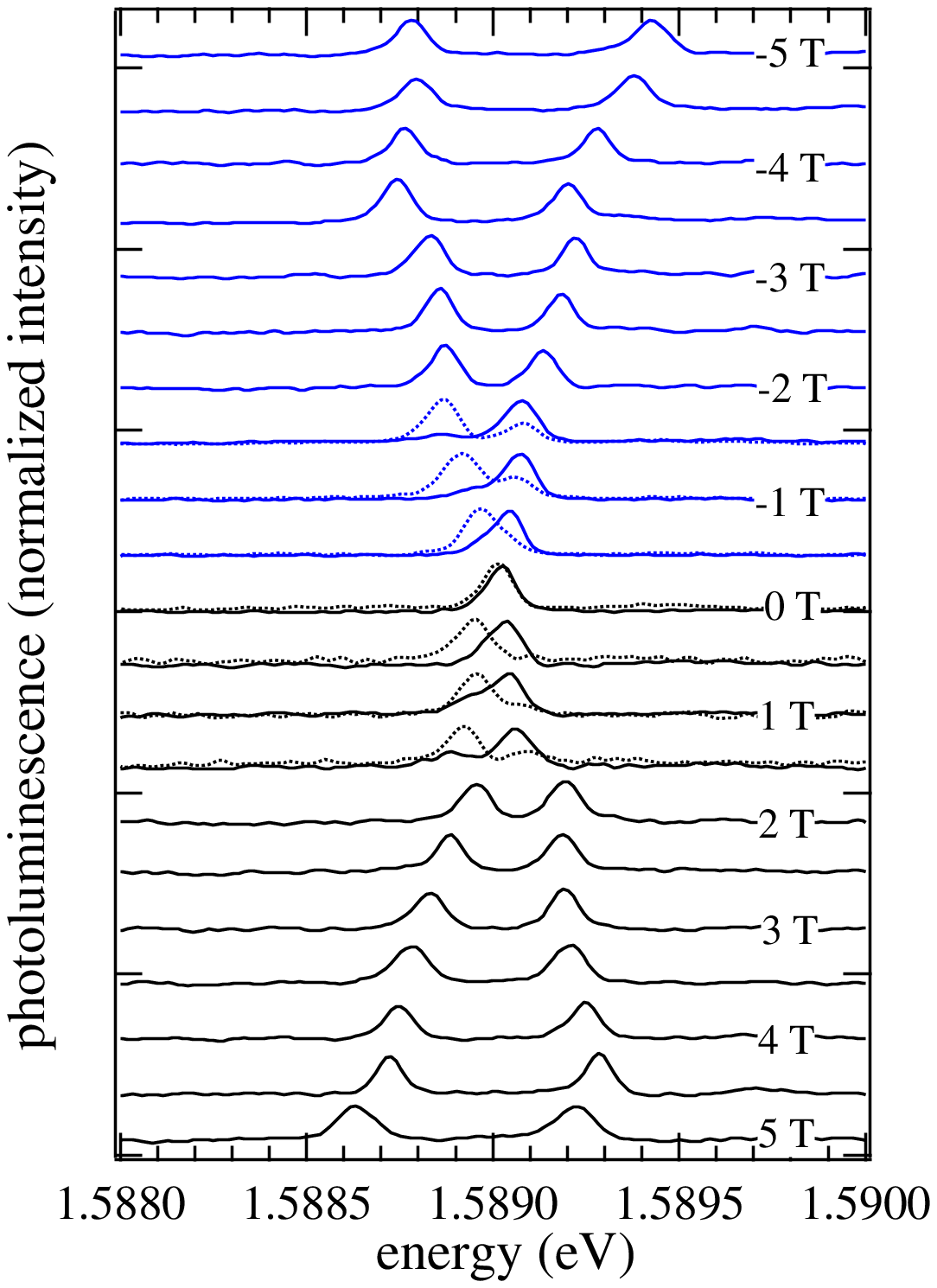}
    \includegraphics[width=200pt]{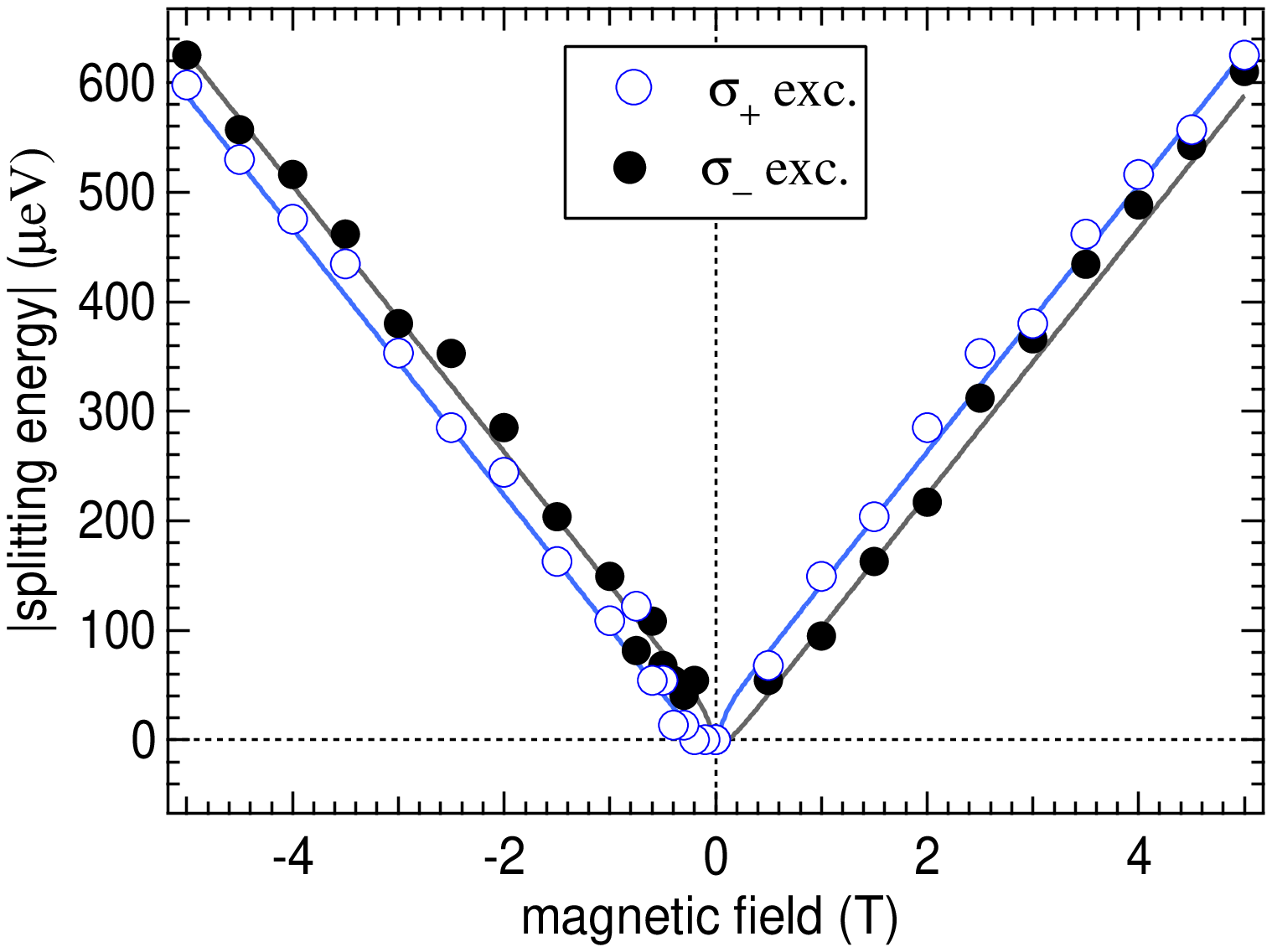}
  \end{center}
  \caption{(a) Excitonic emission energies for $\sigma_+$ excitation. (b) Absolute vales of the splitting energy are plotted for $\sigma_+$ excitation  (open circles) and $\sigma_-$ excitation (black circles).}
    \label{Overhauser}
\end{figure}

Nuclear spin polarization is formed by two-step process in optical pumping. The formation of electron spin polarization is achieved by the circularly-polarized optical excitation in a longitudinal external magnetic field. Next, the electron spin polarization is transferred to the nuclear system via hyperfine interaction, whose Hamiltonian is given by
\begin{equation}
H_{HF}  = {v_0 }\sum\limits_j {A^j } \left| {\psi \left( {{\bf R}_j } \right)} \right|^2 \left( {I_z^j S_z  + \frac{{I_ + ^j S_ -   + I_ - ^j S_ +  }}{2}} \right).
\label{HFeq}
\end{equation} In Eq.~\ref{HFeq}, $v_0$ is the unit cell volume, $A^j$ is the hyperfine constant, $\left| {\psi \left( {{\bf R}_j } \right)} \right|^2$ is the electron density at the $j$th nuclear site ${\bf R}_j$, respectively. 
The interaction consists of two terms; a term proportional to the electronic $S_z$ and nuclear $I_z$ spin polarizations along the direction of external magnetic field and a term including electron and $j$th nuclear raising and lowering operators $S_{+/-}$ and $I^{j}_{+/-}$. The second term describes the dynamic part of hyperfine interaction that is the mutual electron-nuclear spin flips. Through the second term, electron spin polarization is transferred to nuclear spin system.
Then, the formed nuclear spin polarization generates a static effective nuclear magnetic field $B_N$, via the first term, inducing the electronic energy shift. 
The energy shift is known as the Overhauser shift~\cite{OptOrientation}.
The hole in the valence band has the $p$-like wavefunction which vanishes at the position of the nucleus. Thus,  only electron spin polarization contributes to form the nuclear spin polarization. 
The Overhauser field $B_N$ is given by the following;
\begin{equation}
\left\langle {H_{HF} } \right\rangle _N  = A\left\langle {I_z } \right\rangle S_z  = {\mu _B }g_x S_z B_N, 
\end{equation}
where $A$ is the summation of $A^j$ over all the nuclei in a unit cell, and $\left\langle {I_z } \right\rangle $ is average nuclear spin polarization that is determined by the balance of nuclear spin polarization rate and its depolarization rate. 
From the above, 
the Overhauser shift $\Delta E_{OH}$ in QD should be observed as $g_{\rm x} \mu _B B_N$ in the excitonic Zeeman splitting  for circular-polarized excitation. 
  
\begin{figure}[t]
  \begin{center}
    \includegraphics[width=200pt]{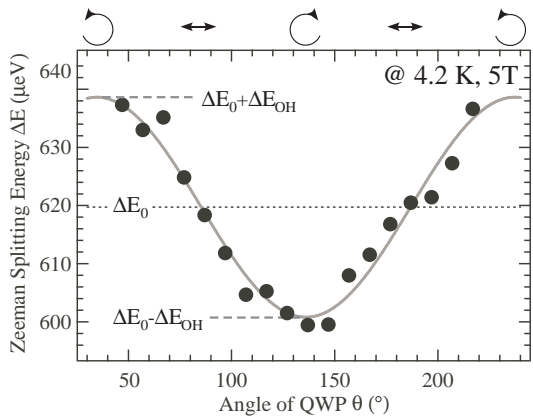}
  \end{center}
  \caption{Energy splitting $\Delta E$ at 5 T is plotted as a function of the angle $\theta$ of the quarterwave plate . The polarization of the excitation light is illustrated above the top axis. In the figure, $\Delta E_0$ is the Zeeman splitting for linearly-polarized excitation that is indicated by the dotted line. The gray curve is a fitting line.}
  \label{Polarization}
\end{figure}
Figure \ref{Overhauser}(a) shows the evolution of the exciton emission in an external magnetic field when exciting a single QD with circularly polarized light. For excitation with $\sigma_+$-polarized light, the energy levels are resolvably split at 0.5 T into two levels. As the magnetic field increases up to 5 T, the splitting of the two levels increases. The similar splitting is seen increasing the magnetic field in opposite direction (up to -5 T), but the splitting energy is found to be slightly different from that in 0-5 T except the range $|B| \le 0.5$ T as shown in Fig. \ref{Overhauser} (b). While this difference is very small (38 $\mu$eV in average), the same result was found in conversion from $\sigma_+$ to $\sigma_-$ excitation under the magnetic field in the same direction. These observations can be explained by the afore-mentioned Overhauser field $B_N$. Under $\sigma_+$ excitation in our system, the induced $B_N$ worked to decrease the external magnetic field $B$ (i.e. $B-B_N$), while $B_N$ worked to increase the external magnetic field in the opposite direction (i.e. $-B-B_N$). Also, the direction of $B_N$ is decided by the direction of the electron spin, which is decided by the polarization of the exciting light. 
In fact, the observed Zeeman splitting $\Delta E$ changed clearly depending on the degree of circular polarization of the excitation light as shown in Fig.~\ref{Polarization}. In Fig.~\ref{Polarization}, $\Delta E$ is plotted as a function of the rotational angle of the quarterwave plate. A very good agreement to $\cos 2(\theta - 45^\circ) $ (solid curve) was obtained as expected for $B_N$. Figure ~\ref{Polarization} also shows the controllability of the $B_N$ by using the degree of circular polarization.
The energy difference in Fig. \ref{Overhauser} (b) is given as $\Delta E_{OH} =g_{b{\rm x}} \mu_B B_N$ by Eq.~\ref{HFeq} and $B_N$ is calculated to be 0.16$\pm$0.01 T for $g_{b{\rm x}}$ of 2.1 and $\Delta E_{OH}$ of 19 $\mu$eV.

\section{Summary}
In summary, we have observed the polarization-dependent shift of the Zeeman splitting in a single InAlAs QD. While the energy difference 19 $\mu$eV and the induced hyperfine field (0.16 T) is small in this InAlAs QD, the magnitude was shown to be controllable by the degree of circular polarization of excitation light.

\acknowledgements{%
This work was partially supported by a Grant-in-Aid for Scientific research in Priority Areas "Semiconductor Nanospintronics" (No.418) of The Ministry of Education, Culture, Sports, Science, and Technology, Japan.

\bibliography{QDDNP.bib}
\end{document}